\documentclass[final]{IEEEtran}
\pdfoutput=1
\usepackage{amsmath, amssymb}
\usepackage{etoolbox}
\usepackage{graphicx}
\usepackage{subfigure}
\usepackage[noadjust]{cite}
\usepackage{color}

\title{Structural Emergency Control Paradigm}

\author{Thanh Long Vu,~\IEEEmembership{Member,~IEEE,} Spyros Chatzivasileiadis,~\IEEEmembership{Member,~IEEE,} Hsiao-Dong Chiang,~\IEEEmembership{Fellow,~IEEE,} and~Konstantin~Turitsyn,~\IEEEmembership{Member,~IEEE}
\thanks{Thanh Long Vu and Konstantin Turitsyn are with the Department of Mechanical Engineering, Massachusetts Institute of Technology, Cambridge, MA, 02139 USA, email: \{longvu, turitsyn\}@mit.edu. Spyros Chatzivasileiadis is with the Department of Electrical Engineering, Technical University of Denmark, email: spchatz@elektro.dtu.dk. Hsiao-Dong Chiang is with School of Electrical and Computer Engineering, Cornell University,  Ithaca, NY, USA, email: chiang@ece.cornell.edu.
}}

\begin{document}

\maketitle
\begin{abstract}
Power grids normally operate at some stable operating condition where power supply and demand are balanced. In response to emergency situations, load shedding is a prevailing approach where local protective devices are activated to cut a suitable amount of load to quickly rebalance the supply demand and hopefully stabilize the system. This traditional emergency control results in interrupted service with severe economic damage to customers. Also, such control is usually less effective due to the lack of coordination among protective devices. In this paper, we propose a novel structural emergency control to render post-fault dynamics from the critical/emergency fault-cleared state to the stable equilibrium point. This is a new control paradigm that does not rely on any continuous measurement or load shedding, as in the classical setup. Instead, the grid is made stable by discretely relocating the equilibrium point and its stability region such that the system is consecutively attracted from the fault-cleared state back to the original equilibrium point. The proposed control is designed by solving linear and convex optimization problems, making it possibly scalable to large-scale power grids. Finally, this emergency control scheme can be implemented by exploiting transmission facilities available on the existing grids.
\end{abstract}


\maketitle

\section{Introduction}

\subsection{Motivation}

The  electric power grid is recognized as the largest engineering achievement in the 20th century. In recent years, it has been experiencing a transformation to an even more complicated
system with an increased number of distributed energy sources and more active and less predictable load endpoints. At the same time, intermittent renewable generation introduces high uncertainty into system operation and may compromise power system stability and security. The existing control operations and modeling approaches, which are largely developed several decades ago for the much more predictable operation of a vertically integrated utility with no fluctuating generation, need to be reassessed and adopted to more stressed operating conditions \cite{camacho2011control, zhao2014design, sharma2015smart,Chen2013139,Xu2016478}. In particular, operating reserves \cite{Wood1996}, traditionally put in place to maintain power system frequency in the presence of uncertainties in production and demand, face limitations in the current grid paradigm. First, the increased uncertainty in production requires new ways of dimensioning the reserves available to the operator at any given moment. Second, because of the substantially higher stochastic component in the current and future power system operation, the power grid becomes increasingly vulnerable to large disturbances, which can eventually  lead to major outages. Such events evolve in time scales much faster than what the secondary or tertiary frequency control can handle.  
Hence, emergency control, i.e., quick actions to recover the stability of a power grid  under critical contingency, is required.

Currently, emergency control of power grids is largely based on remedial actions, special protection schemes (SPS), and load shedding \cite{Vittal2003}, which aim to quickly rebalance power and hopefully stabilize the system. Although these emergency control schemes make the electrical power grid reasonably stable to disturbances, their drawbacks are twofold.
First, some of these emergency actions rely on interrupting electrical service to customers. The unexpected service loss is extremely harmful to customers since it may lead to enormously high economic damage, e.g., it is reported that the economic cost of power interruptions every year in the US is about $\$79$ billion \cite{Lawrence05cost}. Second, protective devices are usually only effective for individual elements, but less effective in preventing the whole grid from collapse. Recent major blackouts exhibit the inability of operators to prevent grid from cascading failures \cite{2003blackout}, regardless of the good performance of individual protective devices. The underlying reason is the lack of coordination among protective devices, which makes them incapable of maintaining the stability of the whole grid. These drawbacks call for  system-level, cost-effective solutions to the emergency control of power grids.

On the other hand, new generations of smart electronic devices provide fast actuation to smart power grids. Advanced transmission resources for active and reactive power flow control
are gradually installed into the system and are expected to be widely available in the future. Besides shunt compensation (switched reactors and capacitors, Static Var Compensators, and STATCOMs), over the last decades a large number of Phase-Shifting Transformers (PSTs) has been installed in power systems all over the world, while a gradually increased installation of Thyristor-Controlled Series Capacitors (TCSCs) has also been observed. Both of these devices can be represented by a variable susceptance (for PST modeling see e.g., \cite{ENTSOE_PST}). At the same time, HVDC lines and HVDC back-to-back converters are installed at several locations, which can also be used for power flow and voltage control.

Motivated by the aforementioned observations, this paper aims to extract more value out of the existing fast-acting controllable grid elements to quickly stabilize the power grid when it is about to lose synchronism after experiencing contingencies (but the voltage is still well-supported). In particular, through the use of PSTs, TCSCs, or HVDC, we propose to adjust selected susceptances and/or power injections in the transmission system to control the post-fault dynamics and thereby stabilize the power system. In the rest of this paper, we will refer to all these devices as FACTS devices. 

One of the most remarkably technical difficulties to realize such a control scheme is that the post-fault dynamics of a power grid possess multiple equilibrium points, each of which has its own stability region (SR), i.e., the set of states from which the post-fault dynamics will converge to the equilibrium point. If the fault-cleared state stays outside the stability region of the stable equilibrium point (SEP), then the post-fault dynamics will result in an unstable condition and eventually, may lead to major failures. 
Real-time direct time-domain simulation,  which exploits advances in computational hardware, can
perform an accurate assessment for post-fault transient dynamics following the contingencies. However, it does not suggest how to properly design the emergency control actions that are guaranteed to drive critical/emergency states back to some stable operating condition.

\begin{figure}[h!]
\centering
\includegraphics[width = 3.2in]{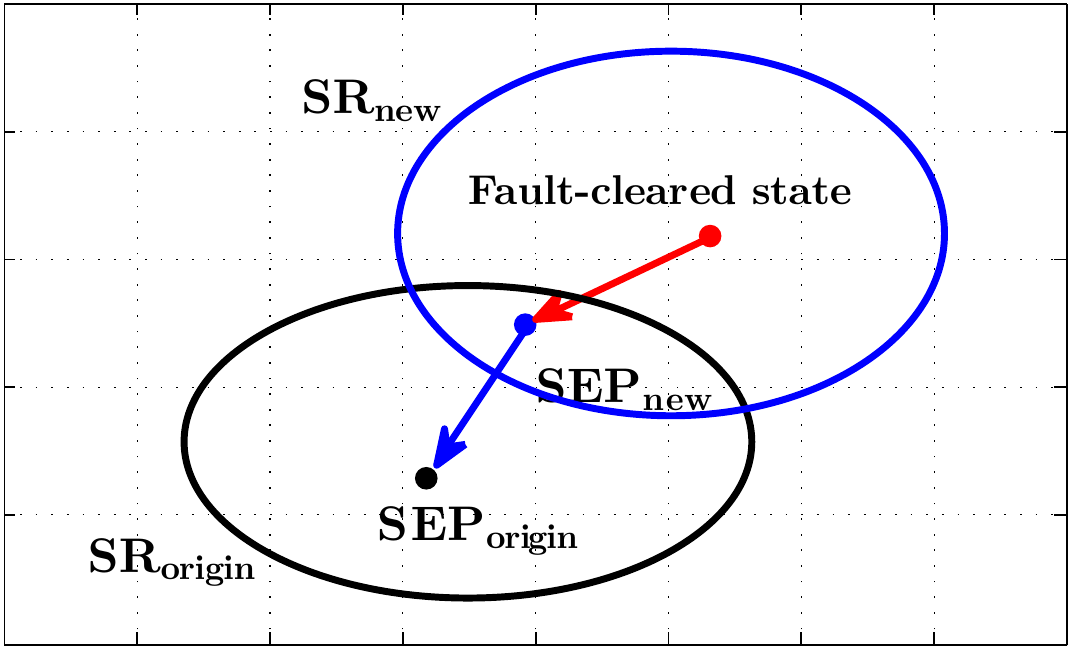}
\caption{Stability-driven smart transmission control: the fault-cleared state is made stable by changing the stable equilibrium point (SEP) through adjusting the susceptances of the network transmission lines.}
\label{fig.EmergencyControl_Idea}
\end{figure}

\subsection{Novelty}

To deal with this technical difficulty, we propose a structural control paradigm to drive post-fault dynamics from critical fault-cleared states to the desired stable equilibrium point. In particular, we will change the transmission line susceptances and/or power injection setpoints to obtain a new stable equilibrium point such that  
the fault-cleared state is guaranteed to stay strictly inside the stability region of this new equilibrium point, as shown in Fig. \ref{fig.EmergencyControl_Idea}. Hence, under the new post-fault dynamics, the system trajectory will converge from the fault-cleared state to the new equilibrium point. If this new equilibrium point stays inside the stability region of the original equilibrium point, then we recover the original line susceptances/power injections and the system state will automatically converge from the new equilibrium point to the original equilibrium point. Otherwise, this convergence  can be performed through a sequence of new transmission control actions which drive the system state to the original equilibrium point through a sequence of other equilibrium points, as shown in Fig. \ref{fig.EmergencyControl_EquilibriumSelection}.

It is worth noting that the proposed control scheme is a new control paradigm which is unusual in classical control systems theory. Indeed, in the proposed control paradigm, we drive the system from the initial state (i.e., the fault-cleared state) to the desired equilibrium point by relocating its equilibrium point and the corresponding stability region. This setup is unusual from the classical control theory point of view where the equilibrium point is usually assumed to be unchanged under the effects of control inputs. 

Compared to the existing control methods, the proposed emergency control method has several advantages, including: 
\begin{itemize}
\item [i)] It belongs to the family of special protection schemes, thus being much faster than secondary and tertiary frequency controls, and making it suitable to handle emergency situations.
 \item [ii)] It avoids load shedding which causes damages and severe economic loss to consumers.
 \item [iii)] The investment for the proposed control is minor since we only employ the  already-installed FACTS devices to change the line impedance/power injection and relocate the equilibrium point.
 \item [iv)] It avoids the usage of continuous measurement of power system state, reducing the resources needed for data storage and processing. The last feature distinguishes the proposed structural control paradigm from other link control methods \cite{PRL.96.164102} where the system state is continuously measured to change the link continuously. 
\end{itemize}

To guarantee the convergence of the post-fault dynamics under control, we utilize our recently introduced Lyapunov function family-based transient stability certificate \cite{VuTuritsyn:2015TAC,VuTuritsyn:2014}. This stability certificate gives us sufficient conditions to assess whether a given post-fault dynamics will converge from a given initial state to a given equilibrium point. In this paper, we construct a new family of Lyapunov functions which are convex and fault-dependent, which can balance the trade-off between computational complexity and conservativeness of the stability certificate. 
Similar idea with such control-Lyapunov function for power systems was also investigated in \cite{ghandhari2001control}, yet this is based on the continuous measurement/control design to change the power injections.

On the practical implementation of the proposed control approach, we note that it may be dangerous if some step went wrong during the whole emergency control procedure, e.g., due to failure of the corresponding FACTS devices. This is at the same degree of risk that the grid would experience in case of protection equipment malfunctions during faults. Operators are familiar with such risks, and there are standardized procedures to ensure the reliable operation of protection relays, e.g., periodic checks, tests, etc. Similar procedures should be followed to ensure the reliable operation of FACTS devices during emergencies.  
In addition, we expect that the proposed approach will act complementary to other emergency control actions. Finally, it is worth noting that in the proposed approach, we only change the susceptances of the transmission lines in the allowable range of FACTS devices, while the selected lines are still on service and the network structure is unchanged. This is different from the line switching approach, which may cause oscillatory behavior after switching action.


The paper is structured as follows.  Section \ref{sec.model}
recalls the structure-preserving model of power
systems and formulates the emergency control problem of
power grids. In Section \ref{sec:LFF}, we construct a new convex, fault-dependent
Lyapunov function family for stability analysis.  
In Section \ref{sec.PostfaultControl}, we design the emergency controls and propose the procedure for
remedial actions. Section \ref{sec.simulations}
numerically illustrates the effectiveness of the proposed emergency control action, and Section \ref{sec.conclusion} concludes the paper.

\section{Network Model and Emergency Control Problem}
\label{sec.model}

\subsection{Network Model}

In this paper, we consider power systems under critical situations when the buses' phasor angles may significantly fluctuate but the buses' voltages are still well-supported and maintained. For such situations, we utilize the standard 
structure-preserving model to describe the dynamics of generators and frequency-dependent dynamic loads in
power systems \cite{bergen1981structure}. This model naturally
incorporates the dynamics of the generators' rotor angle as well as the response of
load power output to frequency deviation. 
Mathematically, the grid is described by an undirected graph
$\mathcal{A}(\mathcal{N},\mathcal{E}),$ where
$\mathcal{N}=\{1,2,\dots,|\mathcal{N}|\}$ is the set of buses and
$\mathcal{E} \subseteq \mathcal{N} \times \mathcal{N}$ is the set
of transmission lines connecting those buses. Here, $|A|$ denotes
the number of elements in set $A.$ The sets of generator buses
and load buses are denoted by $\mathcal{G}$ and $\mathcal{L}$. We assume that the grid is lossless with
constant voltage magnitudes $V_k, k\in \mathcal{N},$ and the
reactive powers are ignored. Then, the structure-preserving model of the system is given by \cite{bergen1981structure}:
\begin{subequations}
\label{eq.structure-preserving}
\begin{align}
\label{eq.structure-preserving1}
 m_k \ddot{\delta_k} + d_k \dot{\delta_k} + \sum_{j \in
  \mathcal{N}_k} a_{kj} \sin(\delta_k-\delta_j) = &P_{m_k},  k \in \mathcal{G},  \\
  \label{eq.structure-preserving2}
  d_k \dot{\delta_k} + \sum_{j \in
  \mathcal{N}_k} a_{kj} \sin(\delta_k-\delta_j) = &-P^0_{d_k},  k \in \mathcal{L},
\end{align}
\end{subequations}
where equation \eqref{eq.structure-preserving1} represents
the dynamics at generator buses and equation
\eqref{eq.structure-preserving2} the dynamics at  load buses.
In these equations, with $k \in \mathcal{G},$ then
 $m_k>0$ is the generator's dimensionless moment of inertia, $d_k>0$ is the term representing primary frequency
controller action on the governor, and $P_{m_k}$ is the input shaft power producing the mechanical torque acting on the rotor of the
$k^{th}$ generator. With $k \in \mathcal{L},$ then $d_k>0$ is  the constant frequency coefficient of load and 
$P^0_{d_k}$ is the nominal load.
Here, $a_{kj}=V_kV_jB_{kj},$ where $B_{kj}$ is the
(normalized)  susceptance of the transmission line $\{k,j\}$ connecting the $k^{th}$ bus and $j^{th}$ bus,
$\mathcal{N}_k$ is the set of neighboring buses of the $k^{th}$
bus.   Note that, the system described by equation \eqref{eq.structure-preserving}
has many stationary points $\delta_k^*$ that are characterized, however,
by the angle differences $\delta_{kj}^*=\delta_k^*-\delta_j^*$
(for a given $P_k$) that solve the following system of power flow-like equations:
\begin{align}
  \label{eq.SEP}
  \sum_{j \in
  \mathcal{N}_k} a_{kj} \sin(\delta_{kj}^*) =P_{k}, k \in \mathcal{N},
\end{align}
where $P_k=P_{m_k}, k \in \mathcal{G},$ and $P_k=-P^0_{d_k}, k \in
\mathcal{L}.$



\subsection{Emergency Control Problem}
\label{sec.formulation}

In normal conditions, a power grid operates at a stable equilibrium point of
the pre-fault dynamics. Under emergency situations,
the system evolves according to the fault-on dynamics laws and moves away
from the pre-fault equilibrium point  to a fault-cleared state $\delta_0$. 
After the fault is cleared, the system evolves according to the post-fault dynamics described by equation \eqref{eq.structure-preserving}. Assume that these post-fault dynamics possess a stable operating condition $\delta^*_{\bf origin}$ with its own stability region. 

The critical situations considered in this paper are when the fault-on trajectory is leaving  polytope $\Pi/2$ defined by inequalities $|\delta_{kj}| \le \pi/2, \forall \{k,j\} \in \mathcal{E},$ i.e., the fault-cleared state $\delta_0$ stays outside  polytope $\Pi/2.$ In normal power systems, protective devices will be activated to disconnect faulted lines/nodes, which will isolate the fault and prevent the post-fault dynamics from instability (this would usually happen at some point beyond a voltage angle difference $\pi/2$).

Avoiding disconnecting line/node, our emergency control objective is to make post-fault dynamics become stable by controlling the post-fault dynamics from
the fault-cleared state $\delta_0$ to the stable equilibrium point $\delta^*_{\bf origin},$ which, e.g., may be an optimum point of some optimal power flow (OPF) problem. 
To achieve this, we consider adjusting the post-fault dynamics through adjusting the susceptance of some selected transmission lines and/or changing power injections. These changes can be implemented by the FACTS devices available on power transmission grids. 
The rationale of this control is based on the observation illustrated in Fig. \ref{fig.EmergencyControl_Idea} that, by appropriately changing the structure of power systems, we can obtain new post-fault dynamics with a new equilibrium point whose region of attraction contains  the fault-cleared state $\delta_0$, and therefore, the new post-fault dynamic is stable. 

Formally, we consider the following control design problem: 
\begin{itemize}
\item [(\textbf{P})] \textbf{Structural Emergency Control Design:} \emph{Given a fault-cleared state $\delta_0$ and the stable equilibrium point $\delta^*_{\bf origin},$ determine the feasible values for susceptances of selected transmission lines and/or feasible power injection such that the post-fault dynamics are driven from the fault-cleared state $\delta_0$  to the original post-fault equilibrium point $\delta^*_{\bf origin}$.}
\end{itemize}

In the next section, we will present the stability certificate for given post-fault dynamics, which will be instrumental in designing a structural emergency control solving problem $(\textbf{P})$ in Section \ref{sec.PostfaultControl}.

\section{Fault-Dependent Transient Stability Certificate}
\label{sec:LFF}

In this section, we recall the Lyapunov function family
approach for transient stability analysis \cite{VuTuritsyn:2014,VuCCT2016}. Then, we construct a new set of fault-dependent Lyapunov functions that are convex and result in an easy-to-verify stability certificate. This set of Lyapunov functions balances the tradeoff between computational tractability and conservativeness of the stability certificate.

\subsection{The Lyapunov Function Family Approach}

In the LFF approach (see \cite{VuTuritsyn:2014,VuCCT2016} for details), the nonlinear couplings and the linear model
are separated, and we obtain an equivalent representation of \eqref{eq.structure-preserving} as
\begin{equation}\label{eq.Bilinear}
 \dot x = A x - B F(C x).
\end{equation}
For the system defined by \eqref{eq.Bilinear}, the LFF approach
proposes to use the Lyapunov functions family  given by:
\begin{align} \label{eq.Lyapunov}
V(x) = \frac{1}{2}x^\top Q x - \sum_{\{k,j\}\in \mathcal{E}}
K_{\{k,j\}} \left(\cos\delta_{kj}
+\delta_{kj}\sin\delta_{kj}^*\right),
\end{align}
in which  the diagonal, nonnegative matrices $K, H$  and the
symmetric, nonnegative matrix $Q$ satisfy the following linear
matrix inequality (LMI):
\begin{align}
\label{eq.QKH}
    \left[   \begin{array}{ccccc}
          A^\top Q+QA  & R \\
          R^\top  & -2H\\
        \end{array}\right] &\le 0,
  \end{align}
with $R = QB-C^\top H-(KCA)^\top$. The classical energy function
is just one element of the large cone of all possible Lyapunov
functions corresponding to a solution of LMI \eqref{eq.QKH}: $Q=\emph{\emph{diag}}(0,...,0,m_1,...,m_m,0,...,0)$,
$K=S,$ and $H=0$. 

Then, we can prove that an estimation for the region of attraction
of the equilibrium point is given by
\begin{align}\label{eq.invariant}
 \mathcal{R_P} = \left\{x \in\mathcal{P}: V(x) < V_{\min}(\mathcal{P})\right\},
\end{align}
where the polytope  $\mathcal{P}$ is
defined by inequalities $|\delta_{kj}+\delta_{kj}^*| \le \pi,
\forall \{k,j\} \in \mathcal{E}$, and $V_{\min}(\mathcal{P})$ is the minimum value of $V(x)$ over the flow-out boundary of polytope $\mathcal{P}$. 
Finally, to determine if the post-fault dynamics are stable, we
check to see if the fault-cleared state $x_0$  is inside the stability
region estimate $\mathcal{R_P}.$

\subsection{The Fault-Dependent Convex Lyapunov Function}
\label{sec:certificate}

A property of the Lyapunov
function $V(x)$ defined in equation \eqref{eq.Lyapunov} is that it may be nonconvex in polytope $\mathcal{P}$, making it computationally complicated to
calculate the minimum value $V_{\min}(\mathcal{P})$. One way to get the convex Lyapunov function is to restrict the state inside the polytope defined by inequalities $|\delta_{kj}|\le \pi/2.$ However, this Lyapunov function can only certify stability for fault-cleared states with phasor differences less than $\pi/2.$

To certify stability for fault-cleared state staying outside polytope $\Pi/2,$ which likely happens in emergency situations, we construct a family of the fault-dependent convex Lyapunov functions.
Assume that the fault-cleared state $x_0$ has a number of phasor differences larger than $\pi/2.$ Usually, this happens when the phasor angle at a node becomes significantly large, making the phasor difference associated with it larger than $\pi/2.$
Without loss of generality, we assume that $|\delta_{ij}(0)|>\pi/2, \forall j\in\mathcal{N}_i$ at some given node $i\in\mathcal{N}$. Also, it still holds that $|\delta_{ij}(0)+\delta_{ij}^*|\le\pi$
for all $j\in\mathcal{N}_i.$ Consider polytope $\mathcal{Q}$ defined by inequalities
\begin{align}
\label{eq.Qmatrix}
|\delta_{ij}+\delta_{ij}^*| &\le\pi, \forall j\in\mathcal{N}_i, \nonumber \\
|\delta_{kj}| &\le\pi/2, \forall j\in\mathcal{N}_k, \forall k \neq i.
\end{align}
Hence, the fault-cleared state is inside polytope $\mathcal{Q}.$ Inside polytope $\mathcal{Q},$ consider the Lyapunov function family \eqref{eq.Lyapunov}
where the matrices $Q,K \ge 0$ satisfying the following LMIs:
\begin{align}
\label{eq.NewQKH}
    \left[   \begin{array}{ccccc}
          A^\top Q+QA  & R \\
          R^\top  & -2H\\
        \end{array}\right] &\le 0, \\
\label{eq.NewQKH1}
Q- \sum_{j\in \mathcal{N}_i}K_{\{i,j\}}C_{\{i,j\}}^\top C_{\{i,j\}} &\ge 0,
  \end{align}
  where $C_{\{i,j\}} $ is the row of matrix $C$ that corresponds to the row containing $K_{\{i,j\}} $ in the diagonal matrix $K.$
From \eqref{eq.Qmatrix} and \eqref{eq.NewQKH1}, we can see that the Hessian of the Lyapunov function inside $\mathcal{Q}$ satisfies
\begin{align}
H(V(x))&=Q + \sum_{\{k,j\}\in \mathcal{E}}K_{\{k,j\}}C_{\{k,j\}}^\top C_{\{k,j\}}\cos\delta_{kj} \nonumber \\&
\ge Q+ \sum_{j\in \mathcal{N}_i}K_{\{i,j\}}C_{\{i,j\}}^\top C_{\{i,j\}}\cos \delta_{ij} \nonumber \\&
\ge Q- \sum_{j\in \mathcal{N}_i}K_{\{i,j\}}C_{\{i,j\}}^\top C_{\{i,j\}} \ge 0.
\end{align}
As such, the Lyapunov function is convex inside polytope $\mathcal{Q}$
and thus, the corresponding minimum value $V_{\min}(\mathcal{Q}),$ defined over the flow-out boundary of $\mathcal{Q},$ can be calculated in polynomial time.
Also, the corresponding estimate for region of attraction is
given by
\begin{align}\label{eq.RoAestimate}
 \mathcal{R_Q} = \left\{x \in\mathcal{Q}: V(x) < V_{\min}\right\},
\end{align}
with
\begin{align}
\label{eq.Vmin2} V_{\min}=V_{\min}(\mathcal{Q})=\mathop {\min}\limits_{x \in
\partial\mathcal{Q}^{out}} V(x).
\end{align}

The convexity of $V(x)$ in polytope $\mathcal{Q}$ allows us to quickly compute the minimum value $V_{\min}$ and come up with an easy-to-verify stability certificate. Therefore, by exploiting properties of the fault-cleared state, we have a family of fault-dependent Lyapunov functions that balance the tradeoff between computational complexity and conservativeness. It is worth noting that though the Lyapunov function is fault-dependent, we only need information for the fault-cleared states instead of the full fault-on dynamics.

Another point we should note is that LMIs \eqref{eq.NewQKH}-\eqref{eq.NewQKH1} provide us with a family of Lyapunov functions guaranteeing the stability of the post-fault dynamics. For a given fault-cleared state, we can find the best suitable function in this family to certify its stability. The adaptation algorithm is similar to that in \cite{VuTuritsyn:2014}, with the only difference being the augment of inequality \eqref{eq.NewQKH1}, i.e., $Q- \sum_{j\in \mathcal{N}_i}K_{\{i,j\}}C_{\{i,j\}}^\top C_{\{i,j\}} \ge 0.$ More details can be found in Appendix \ref{appendix}.

\section{Structural Emergency Control Design}
\label{sec.PostfaultControl}

\begin{figure}[t!]
\centering
\includegraphics[width = 3.2in]{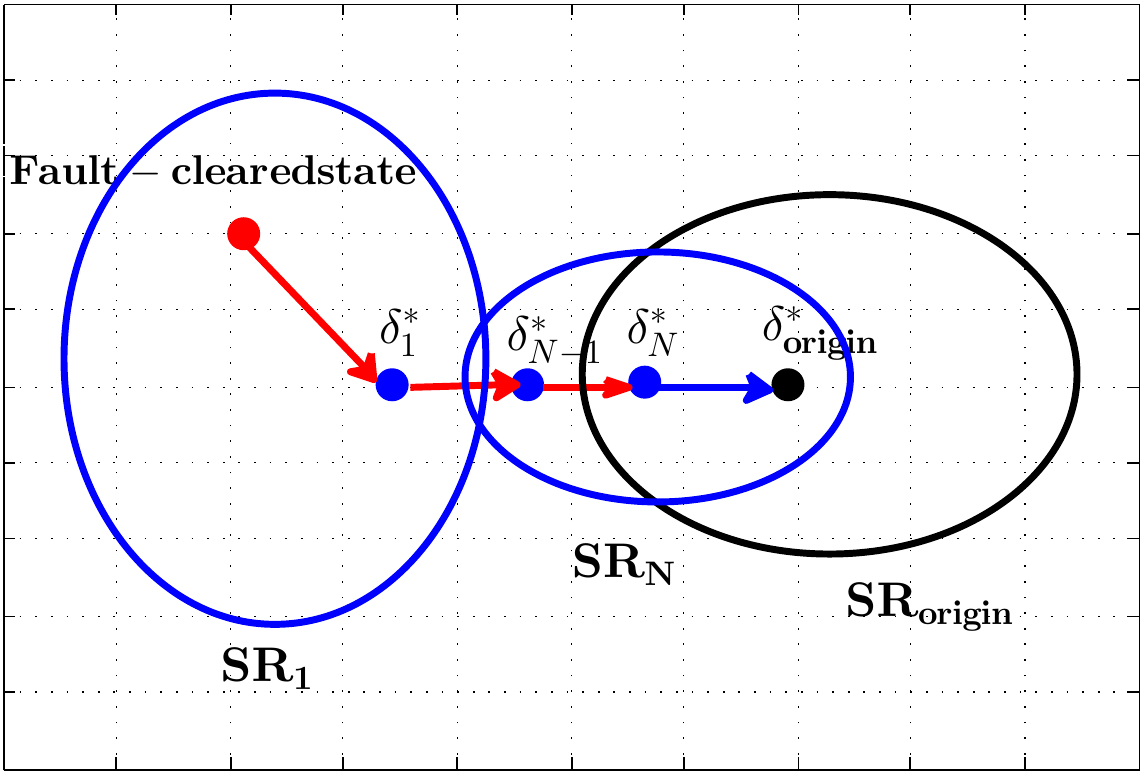}
\caption{Selection of the sequence of stable equilibrium points $\delta^*_{i}, i=1,...,N,$ such that the fault-cleared state
is driven through the sequence of equilibrium points back to the original equilibrium point $\delta^*_{\bf origin}$.}
\label{fig.EmergencyControl_EquilibriumSelection}
\end{figure}

In this section, we solve the post-fault emergency control problem $\textbf{(P)}.$ As illustrated in Fig. \ref{fig.EmergencyControl_EquilibriumSelection}, to render the post-fault dynamics from the fault-cleared state $x_0$ to the equilibrium point $\delta^*_{\bf origin},$ we will find a sequence of stable equilibrium points $\delta^*_1,...,\delta^*_N$ with their corresponding region of attractions ${\bf SR_1,...,SR_N}$ such that
\begin{align}
x_0 \in {\bf SR_1}, \delta_1^*\in {\bf SR_2},..., \delta_{N-1}^*\in {\bf SR_{N}},\delta_N^* \in {\bf SR_{origin}}.
\end{align}
Then, the post-fault dynamics can be attracted from the fault-cleared state $x_0$ to the original equilibrium point $\delta^*_{\bf origin}$ through a sequence
of appropriate structural changes in the power network. In this section, we will show that we only need to determine a finite number of equilibrium points through solving convex optimization problems.

Recall that, the equilibrium point $\delta^*$ is a solution to the power flow-like equations:
 \begin{align}
 \label{eq.PFE}
 \sum_{j\in \mathcal{N}_k}V_kV_jB_{kj}\sin \delta^*_{{kj}}=P_k, \forall k \in \mathcal{N}.
 \end{align}
 As such, the sequence of equilibrium points $\delta^*_1,...,\delta^*_N$ can be obtained by appropriately changing the susceptances $\{B_{kj}\}$ of the transmission lines or by changing the power injection $P_k$.
 
 In the following, we will design the first equilibrium point $\delta^*_1$ by changing the selected line susceptances/power injection, and then design the other equilibrium points $\delta^*_2,...,\delta^*_N$ by only adjusting the
 susceptances of selected transmission lines. We note that, in each control step,  the susceptances of transmission lines or the power injections will only be changed one time. This scheme eliminates the need for the continuous measurement and continuous control actuation required in traditional feedback control practices.

Designing the first equilibrium point $\delta^*_1$ to drive the system from an unstable state (i.e., the fault-cleared state $x_0$) to the stable state $\delta_1^*$ will be performed in a way that differs from designing the other equilibrium points which serve to drive the system from the stable state $\delta_1^*$ to the original stable state $\delta^*_{\bf origin}.$ 

\subsection{Design the first equilibrium point $\delta^*_1$ by changing the transmission susceptances}

We need to find the new susceptances of transmission lines such that the equilibrium point $\delta_1^*$, which has the stability region ${\bf SR_1}$, contains $x_0.$
Consider the energy function in the Lyapunov function family \eqref{eq.Lyapunov}:
\begin{align}
\label{eq.energy}
    V(x) &=\sum_{k \in \mathcal{N}}\frac{m_k \dot{\delta}_k^2}{2} - \sum_{\{k,j\}\in \mathcal{E}}B_{kj}V_kV_j(\cos \delta_{kj}+\delta_{kj}\sin\delta_{1_{kj}}^*)\nonumber\\& 
    =\sum_{k \in \mathcal{N}}\frac{m_k \dot{\delta}_k^2}{2} - \sum_{\{k,j\}\in \mathcal{E}}B_{kj}V_kV_j\cos \delta_{kj} -\sum_{k\in\mathcal{N}}P_k\delta_k.
\end{align}
We will find $\{B_{kj}\}$ such that $x_0 \in \mathcal{R}_{\mathcal{Q}}(\delta_1^*),$ i.e., $x_0\in \mathcal{Q}$ and $V(x_0)<V_{\min}.$ Note that, $V(x_0)$ is a linear function of $\{B_{kj}\}.$ Generally, $V_{\min}$ is a nonlinear function of $\{B_{kj}\}$. However, if we use the lower bound of $V_{\min}$ \cite{VuTuritsyn:2014}, we can have a bound $V_{\min}^{lower}$ that is linear in $\{B_{kj}\}.$ Then, the condition $V(x_0)<V_{\min}^{lower}$ is a linear matrix inequality, and thus can be solved quickly by convex optimization solvers to obtain a feasible solution of $V(x_0)<V_{\min}$. 

\subsection{Design the first equilibrium point $\delta_1^*$ by changing the power injections}

Another way to design $\delta^*_1$ is by changing the power injection. The post-fault dynamics are locally stable when the equilibrium point stays inside the polytope defined by the inequalities $|\delta_{kj}|<\pi/2$ \cite{Dorfler:2013}. To make the post-fault dynamics stable, we can place the equilibrium point
far away from the margin $|\delta_{kj}|=\pi/2,$ i.e., making the phasor differences $\delta_{kj}$ near $0.$ As such, to search for the equilibrium point $\delta^*_1$ such that $x_0\in {\bf SR_1},$ we will find the equilibrium point $\delta^*_1$ such that its phasor differences are as small in magnitude as possible.  

We recall in \cite{Dorfler:2013} that, for almost all power systems, to make sure $|\delta^*_{kj}|<\gamma <\pi/2$, we need
\begin{align}
\label{eq.SynchronizationCondition}
\|L^{\dag}p\|_{\mathcal{E},\infty} \le \sin\gamma.
\end{align}
Here, $L^\dag$ is the pseudoinverse of the network Laplacian
matrix, $p=[P_1,...,P_{|\mathcal{N}|}]^\top,$ and
$\|x\|_{\mathcal{E},\infty}=\max_{\{i,j\}\in
\mathcal{E}}|x(i)-x(j)|.$ 
Therefore, to make the phasor differences of  the equilibrium point $\delta^*_1$ as small as possible, we will find the power injection $P_k$ such that
$\|L^{\dag}p\|_{\mathcal{E},\infty}$ as small as possible, i.e., minimizing $\|L^{\dag}p\|_{\mathcal{E},\infty}.$ 
Note that, with fixed susceptances, the Laplacian matrix $L^\dag$ is fixed. As such, minimizing $\|L^{\dag}p\|_{\mathcal{E},\infty}$
over all possible power injections is a linear optimization problem.

After designing the first equilibrium point $\delta^*_1,$  we can check if $x_0 \in {\bf SR_1}$ by applying the stability certificate presented in the previous section. In particular, given the equilibrium point $\delta^*_1$ and the fault-cleared state $x_0,$ we can adapt the Lyapunov function family to find a suitable function $V(x)$ such that
$V(x_0)<V_{\min}.$ A similar adaptation algorithm with what was introduced in \cite{VuTuritsyn:2014} can find such a Lyapunov function after a finite number of steps.

We summarize the procedure as follows.

{\bf Procedure 1.}

\begin{itemize}
    \item Minimize the linear function $\|L^{\dag}p\|_{\mathcal{E},\infty}$ over the power injection space;
    \item Calculate the new equilibrium point from the optimum value of the power injection;
    \item Given the new equilibrium point, utilize the adaptation algorithm to search for a Lyapunov function that can certify stability for the fault-cleared state $x_0.$
\end{itemize}

\subsection{Design the other equilibrium points by changing the susceptances of transmission lines}

Now, given the equilibrium points $\delta^*_1$ and $\delta^*_{\bf origin},$ we will design a sequence of stable equilibrium points $\delta^*_2,...,\delta^*_N$ such that
$\delta_1^*\in {\bf SR_2},..., \delta_{N-1}^*\in {\bf SR_{N}},\delta^*_N \in {\bf SR_{origin}}.$ Since all of these stable equilibrium points stay inside  polytope $\Pi/2,$ this design can be feasible. 

{\bf Case 1:} The number of transmission lines that we can change is larger than the number of buses $|\mathcal{N}|$ (i.e., the number of lines with FACTS/PST devices available is larger than $|\mathcal{N}|$), and there are no constraints on the corresponding susceptances. Then, given the equilibrium point $\delta^*,$ it is possible to solve equation \eqref{eq.PFE} with variables the varying susceptances. Now, we can choose the sequence of stable equilibrium points $\delta^*_2,...,\delta^*_N$ equi-spaced between the equilibrium points $\delta^*_1$ and $\delta^*_{\bf origin},$ and find the corresponding susceptances. Then we use the stability certificate presented in Section III to check if $\delta_1^*\in {\bf SR_2},..., \delta_{N-1}^*\in {\bf SR_{N}},\delta^*_N \in {\bf SR_{origin}}.$

{\bf Case 2:} The number of transmission lines that we can change is smaller than the number of buses $|\mathcal{N}|,$ or there are some  constraints on the corresponding susceptances. Then, it is not always possible to find the suitable susceptances satisfying equation \eqref{eq.PFE} from the given equilibrium point $\delta^*.$ 

\begin{figure}[t!]
\centering
\includegraphics[width = 3.2in]{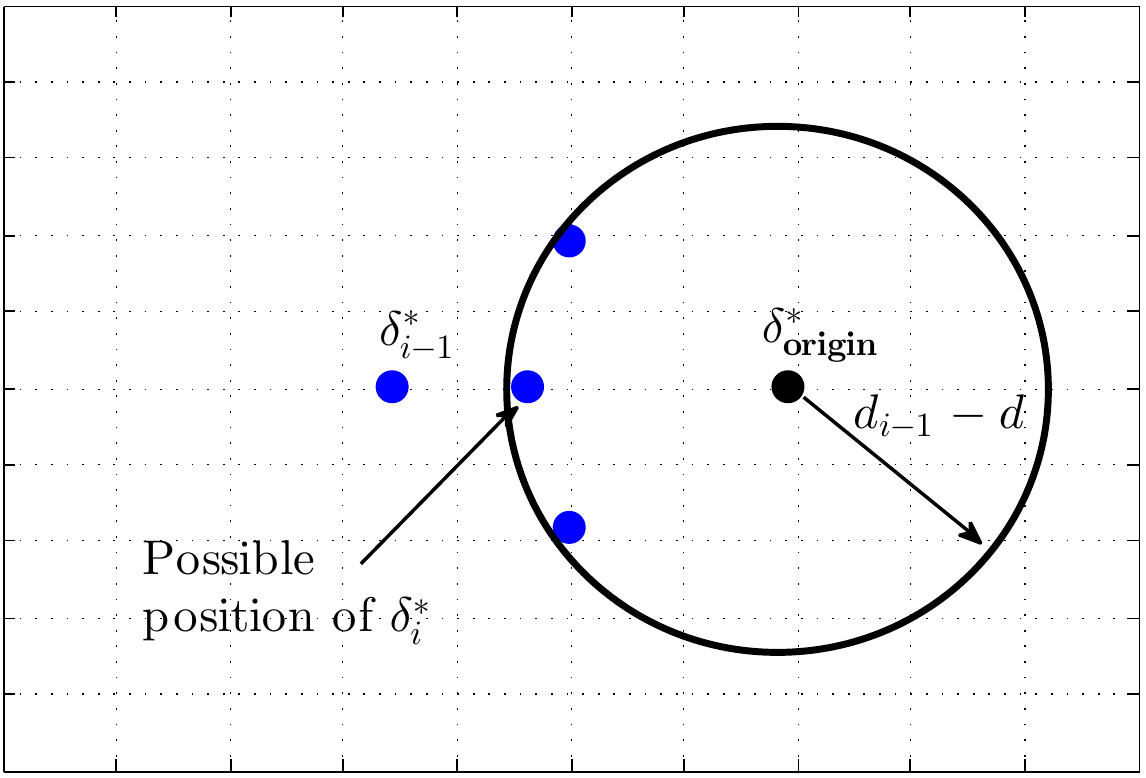}
\caption{Localization of $\delta^*_{i}$ as the closest point to $\delta_{i-1}^*$ that stays inside the ball around $\delta_{\bf origin}^*$ with the radius
$d_{i-1}(\delta_{i-1}^*,\delta_{\bf origin}^*)-d$. The minimization of the distance is taken over all the reachable susceptance values of the selected transmission lines. Here,  minimizing the distance between $\delta^*_{i}$ and $\delta^*_{i-1}$ enables the convergence from $\delta^*_{i-1}$ to $\delta^*_{i}.$ The constraint that $\delta^*_{i}$ stays in the ball will make sure that the distance from the designed equilibrium point to $\delta_{\bf origin}^*$ is decreasing, and eventually, the equilibrium point stays closed enough to $\delta_{\bf origin}^*$ such that the system will converge from this equilibrium point to $\delta_{\bf origin}^*$.}
\label{fig.EmergencyControl_EquilibriumLocalization}
\end{figure}

In each step, to allow the convergence from $\delta_{i-1}^*$ to $\delta_i^*,$ we will search over all the reachable susceptance values of selected  transmission lines the best one that minimizes the distance from $\delta_{i-1}^*$ to $\delta_i^*$. At the same time, we will make the distance
from these equilibrium points to the original equilibrium point $\delta^*_{\bf origin}$ strictly decreasing to make sure that we only need to design a finite number of equilibrium points. Intuitively, the localization of the equilibrium point $\delta^*_{i}$ is shown in Fig. \ref{fig.EmergencyControl_EquilibriumLocalization}. Accordingly, for the reachable set of transmission susceptances, we define $\delta^*_2$ as the closest possible equilibrium point   to $\delta^*_1$ and the distance between $\delta_2^*$ and $\delta_{\bf origin}^*$ satisfies 
\begin{align}
\label{eq.DistanceCondition}
d_2(\delta^*_2,\delta^*_{\bf origin})\le d_1(\delta_1^*,\delta^*_{\bf origin})-d,\end{align}
where $d>0$ is a constant. Similarly, $\delta^*_3$ is the closest possible equilibrium point  to $\delta^*_2,$ and satisfies
\begin{align}
\label{eq.DistanceCondition1}d_3(\delta^*_3,\delta^*_{\bf origin}) \le d_2(\delta_2^*,\delta^*_{\bf origin})-d.\end{align},
and so on.
Here, $d>0$ is a sufficiently small constant chosen such that the convergence from $\delta_{i-1}^*$ to $\delta_{i}^*$ is satisfied for all $i=2,...,N$,
and $d_i(\delta^*_i,\delta)$ is the distance from $\delta$ to the equilibrium point $\delta_i^*$, which is defined via $\{B_{kj}^{(i)}\},$ i.e.,
\begin{align*}
d_i(\delta^*_i,\delta) &= \sum_{k \in \mathcal{N}} \big(\sum_{j\in \mathcal{N}_k}V_kV_jB_{kj}^{(i)}(\sin\delta_{i_{kj}}^*-\sin\delta_{{kj}})\big)^2 \nonumber \\&
= \sum_{k \in \mathcal{N}} \big(P_k-\sum_{j\in \mathcal{N}_k}V_kV_jB_{kj}^{(i)}\sin\delta_{{kj}}\big)^2.
\end{align*}
Note that, with $d=0,$ the trivial solution to all of the above optimization problems is $\delta_N^* \equiv ... \equiv \delta_2^* \equiv \delta_1^*,$
and the convergence from $\delta_{i-1}^*$ to $\delta_{i}^*$ is automatically satisfied.
Nonetheless, since each of the equilibrium points has a nontrivial stability region, there exists a sufficiently small $d>0$ such that
the convergence from $\delta_{i-1}^*$ to $\delta_{i}^*$ must still be satisfied for all $i=2,...,N.$ 

On the other hand, since $d_i(\delta^*_i,\delta^*)$ is a quadratic function of $\{B_{kj}^{(i)}\},$ 
defining $\delta^*_2,...,\delta^*_N$ can be described by the quadratically
constrained
quadratic
program (QCQP) in $\{B_{kj}^{(i)}\}:$
\begin{align}
\label{eq.DefiningEquilibrium}
    &\min_{\{B^{(i)}_{kj}\}} d_i(\delta^*_i,\delta^*_{i-1}) \\
    {\bf s.t.\;\; } & d_i(\delta^*_i,\delta^*_{\bf origin}) \le d_{i-1}(\delta^*_{i-1},\delta^*_{\bf origin})-d \nonumber.
\end{align}
In optimization problem \eqref{eq.DefiningEquilibrium}, $d_{i-1}(\delta^*_{i-1},\delta^*_{\bf origin})$ is a constant obtained from the previous step.
Note that, the condition $d_i(\delta^*_i,\delta^*_{\bf origin}) \le d_{i-1}(\delta^*_{i-1},\delta^*_{\bf origin})-d$ will probably place $\delta^*_i$
between $\delta^*_{i-1}$ and $\delta^*_{\bf origin},$ which will automatically guarantee that $\delta^*_i$ stays inside  polytope $\Pi/2$.  Also, since the equilibrium points are strictly staying inside polytope $\Pi/2,$ the objective function $d_i(\delta^*_i,\delta^*_{i-1}) $ and the constraint function
$d_i(\delta^*_i,\delta^*_{\bf origin})$ are strictly convex functions of $\{B_{kj}^{(i)}\}.$ As such,  QCQP \eqref{eq.DefiningEquilibrium}  is convex and can be quickly solved using convex optimization solvers.

When all of these optimization problems are feasible, then with $d>0$ from Eqs. \eqref{eq.DistanceCondition}-\eqref{eq.DistanceCondition1}, we have
\begin{align}
d_1(\delta_1^*,\delta^*_{\bf origin})& \ge d_2(\delta_2^*,\delta^*_{\bf origin})+d \ge...\nonumber \\&\ge d_N(\delta_N^*,\delta^*_{\bf origin})+(N-1)d 
\nonumber \\&\ge (N-1)d. 
\end{align}
As such, $N\le 1+ (d_1(\delta_1^*,\delta^*_{\bf origin})/d),$ and hence, there is only a finite number of equilibrium points $\delta_2^*,...,\delta_N^*$ that
we need to determine.

\subsection{Structural remedial actions}

We propose the following procedure of emergency controls to render post-fault dynamics from critical fault-cleared states to the desired stable equilibrium point.

\begin{itemize}

\item {\bf Initialization:} Check if the given fault-cleared state $\delta_0$ stays inside the stability region of the original equilibrium point $\delta^*_{\bf origin}$ by utilizing the stability certificate in Section \ref{sec:certificate}. If not, go to {\bf Step 1}, otherwise end.

\item {\bf Step 1:} Fix the susceptances  and change the power injection such that the fault-cleared state $\delta_0$ stays inside the stability region ${\bf SR_1}$ of the new equilibrium point $\delta^*_1.$ The post-fault dynamics with power injection control will converge from the fault-cleared state $\delta_0$ to the equilibrium point $\delta_1^*.$ Recover the power injections after the post-fault dynamics converge to $\delta_1^*.$

Check whether $\delta^*_1$ stays in the stability region of the original equilibrium point $\delta^*_{\bf origin}$
by using the Lyapunov function stability certificate. If this holds true, then the post-fault dynamics will converge from the new equilibrium point to the original equilibrium point. If not, then go to Iterative Steps.

\item {\bf Iterative Steps:}   Determine the transmission susceptances such that the sequence of stable equilibrium points $\delta^*_2,...,\delta^*_N$ satisfies that $\delta_1^*\in {\bf SR_2},..., \delta_{N-1}^*\in {\bf SR_{N}},\delta^*_N \in {\bf SR_{origin}}.$ 
Apply consecutively the susceptance changes on the transmission lines to render the post-fault dynamics from $\delta^*_1$ to $\delta^*_N.$

\item {\bf Final Step:} Restore the susceptances to the original susceptances. Then, the post-fault dynamics will automatically converge from $\delta^*_N$ to the original equilibrium point $\delta^*_{\bf origin}$
since $\delta^*_N \in {\bf SR_{origin}}.$

\end{itemize}


\section{Numerical Validation}
\label{sec.simulations}

\subsection{Kundur 9-Bus 3-Generator System}
\begin{figure}[t!]
\centering
\includegraphics[width = 3.2in]{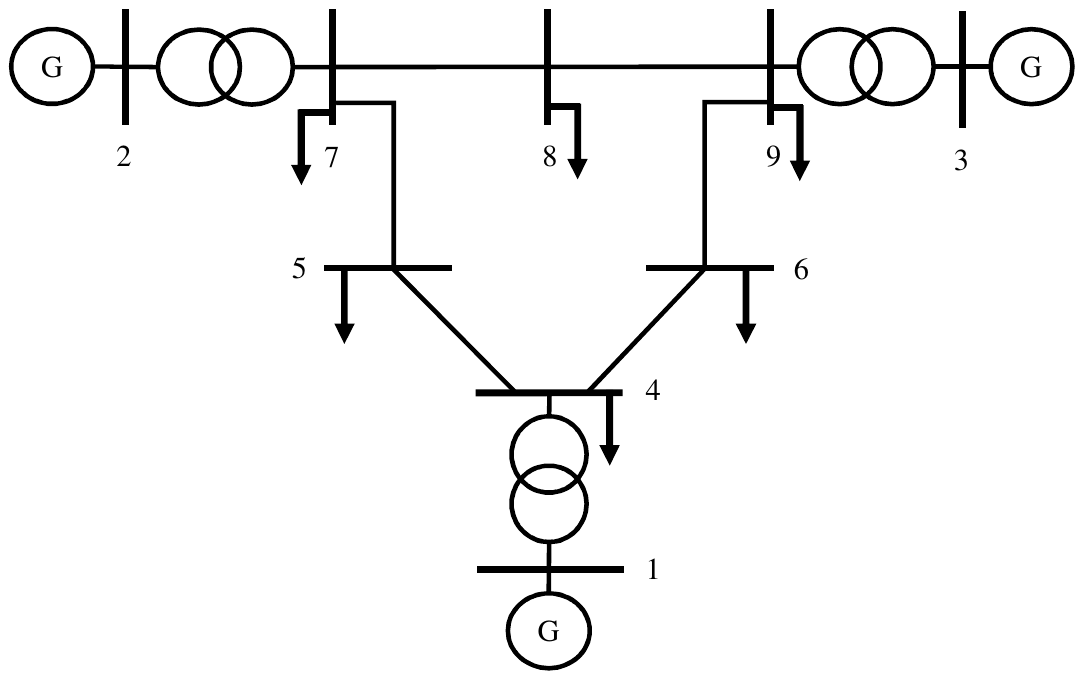}
\caption{A 3 generator 9 bus system with frequency-dependent dynamic
loads.} \label{fig.3generator9bus}
\end{figure}

Consider the 9-bus 3-machine system depicted in Fig.
\ref{fig.3generator9bus}  with 3 generator buses and 6 load buses.
The susceptances of the transmission lines are as follows
\cite{Anderson:2003}:
$B_{14}=17.3611 p.u.,B_{27}=16.0000 p.u.,B_{39}= 17.0648 p.u.,
B_{45}=11.7647 p.u., B_{57}= 6.2112p.u., B_{64}=10.8696p.u.,
B_{78}= 13.8889p.u.,B_{89}=9.9206p.u., B_{96}=5.8824p.u.$ 
    The parameters for generators are $m_1=0.1254, m_2=0.034, m_3=0.016, d_1=0.0627, d_2=0.017, d_3=0.008.$ For simplicity, we take $d_k=0.05, k=4\dots,9.$ 



    Assume that the fault trips the line between buses $5$ and $7$ and make the power injection variate. When the fault is cleared this line is re-closed. We also assume the fluctuation of the generation (probably due to renewables) and load such that the bus
voltages $V_k$, mechanical inputs $P_{m_k}$, and steady state load
$-P_{d_k}^0$ of the post-fault dynamics after clearing the fault are given in Tab. \ref{tab.data9bus}. The stable
operating condition is calculated 
 as $\delta_{\bf origin}^*=[-0.1629\;
    0.4416\;
    0.3623\;
   -0.3563\;
   -0.3608\;
   -0.3651\;
    0.1680\;
    0.1362\;
    0.1371]^\top, \dot{\delta}_{\bf origin}^*=0.$ However,  the fault-cleared state, with angles
   $\delta_0=[0.025 \;-0.023\; 0.041\; 0.012\; -2.917\; -0.004\; 0.907\; 0.021\; 0.023]^\top$ and generators angular velocity $[-0.016\; -0.021\; 0.014]^\top,$  stays outside polytope $\Pi/2.$ By our adaptation algorithm, we do not find a suitable Lyapunov function certifying the convergence of this fault-cleared state to the original equilibrium point $\delta_{\bf origin}^*,$ so this fault-cleared state may be unstable. We will design emergency control actions to bring the
    post-fault dynamics from the possibly unstable fault-cleared state to the equilibrium point $\delta_{\bf origin}^*.$ All the convex optimization problems associated in the design will be solved by CVX software.
    
    \begin{figure}[t!]
\centering
\includegraphics[width = 3.2in]{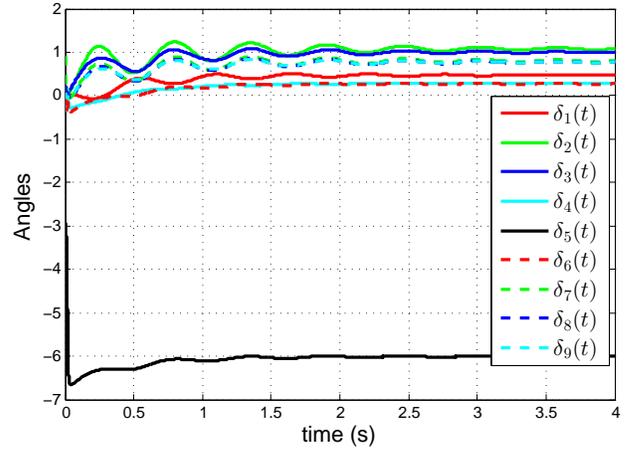}
\caption{Buses angular dynamics when the proposed control is not employed} \label{fig.NoControl_Angle}
\end{figure}
    
\begin{table}[ht!]
\centering
\begin{tabular}{|c|c|c|}
  \hline
  Node & V (p.u.) & $P_k$ (p.u.) \\
  \hline
  1 & 1.0284 & 3.6466 \\
  2 & 1.0085 & 4.5735 \\
  3 & 0.9522 &  3.8173 \\
  4 & 1.0627 & -3.4771 \\
  5 & 1.0707 & -3.5798 \\
  6 & 1.0749 & -3.3112 \\
  7 & 1.0490 & -0.5639 \\
  8 & 1.0579 &  -0.5000 \\
  9 & 1.0521 &  -0.6054 \\
  \hline
\end{tabular}
\caption{Bus voltages, mechanical inputs, and static
loads.}\label{tab.data9bus}
\end{table}

\subsubsection{Designing the first equilibrium point}

Assume that the three generators 1-3 are dispatchable and terminal loads at buses 4-6 are controllable, while terminal loads at the other buses are fixed. We design the first equilibrium point by changing the power injections of the three generators 1-3 and load buses 4-6. With the original power injection, 
$\|L^{\dag}p\|_{\mathcal{E},\infty}=0.5288.$ Using CVX software to minimize $\|L^{\dag}p\|_{\mathcal{E},\infty},$ we obtain the new power injections at buses 1-6 as follows: $P_1= 0.5890, P_2=
    0.5930, P_3=
    0.5989, P_4=
   -0.0333, P_5=
   -0.0617,$ and $P_6=
   -0.0165.$ Accordingly,
the minimum value of $\|L^{\dag}p\|_{\mathcal{E},\infty}=0.0350 < \emph{\emph{sin}}(\pi/89).$ Hence, the first equilibrium point obtained from equation \eqref{eq.SEP} will stay in the polytope defined by the inequalities $|\delta_{kj}|\le \pi/89, \forall \{k,j\}\in\mathcal{E},$ and can be approximated by
$\delta^*_1 \approx L^{\dag}p=[  0.0581\;
    0.0042\;
    0.0070\;
    0.0271\;
    0.0042\;
    0.0070\;
   -0.0308\;
   -0.0486\;
   -0.0281]^\top$. 
\begin{figure}[t!]
\centering
\includegraphics[width = 3.2in]{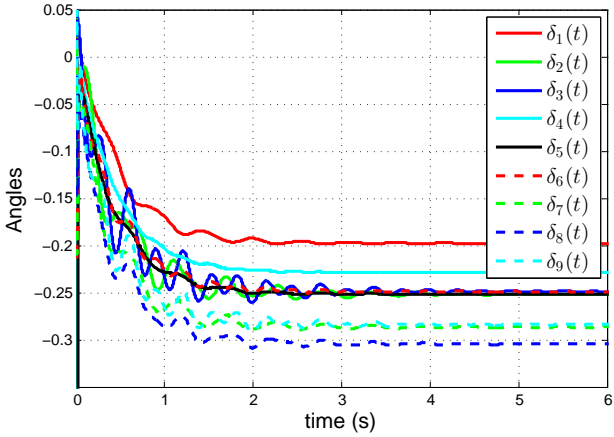}
\caption{Effect of power injection control: Convergence of buses angles from the fault-cleared state to $\delta_1^*$ in the post-fault dynamics} \label{fig.InjectionControl_Angle}
\end{figure}
\begin{figure}[t!]
\centering
\includegraphics[width = 3.2in]{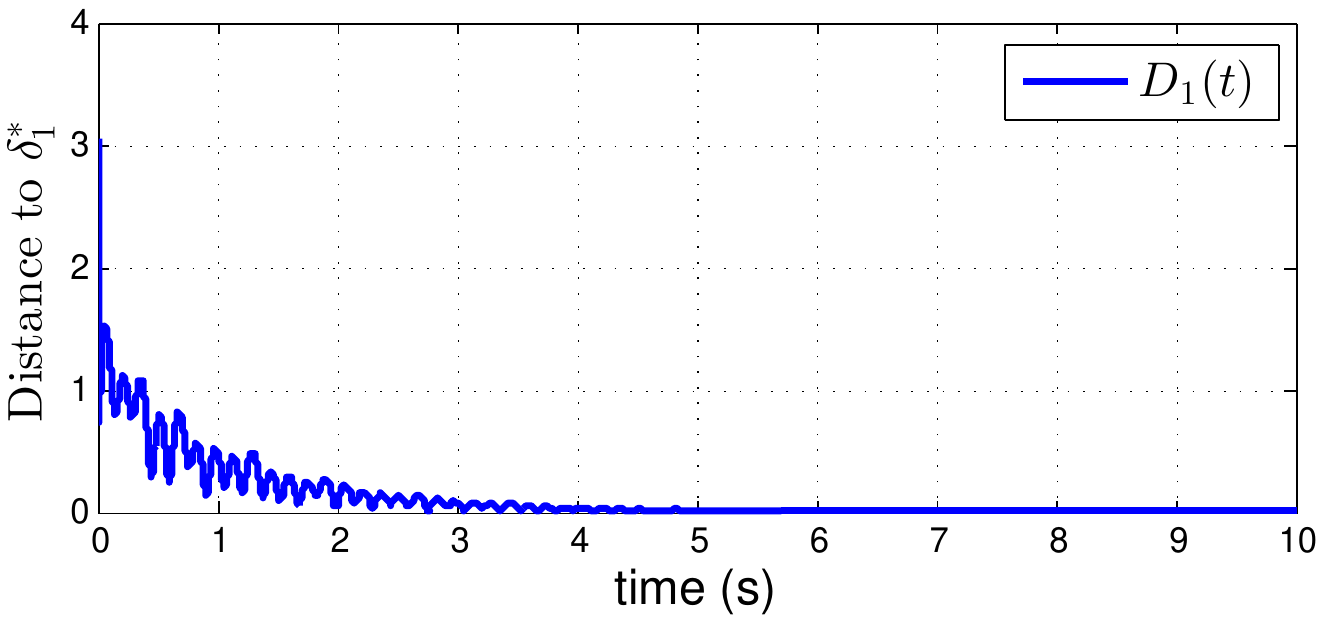}
\caption{Effect of injection control: the convergence of the distance $D_1(t)$ to $0$. Here, the Euclid distance $D_1(t)$ between a post-fault state and the first equilibrium point $\delta_1^*$
is defined as $D_1(t)=\sqrt{\sum_{i=2}^{9} (\delta_{i1}(t)-\delta_{1_{i1}}^*)^2}$.} 
\label{fig.InjectionControl_Distance}
\end{figure}
\begin{figure}[t!]
\centering
\includegraphics[width = 3.2in]{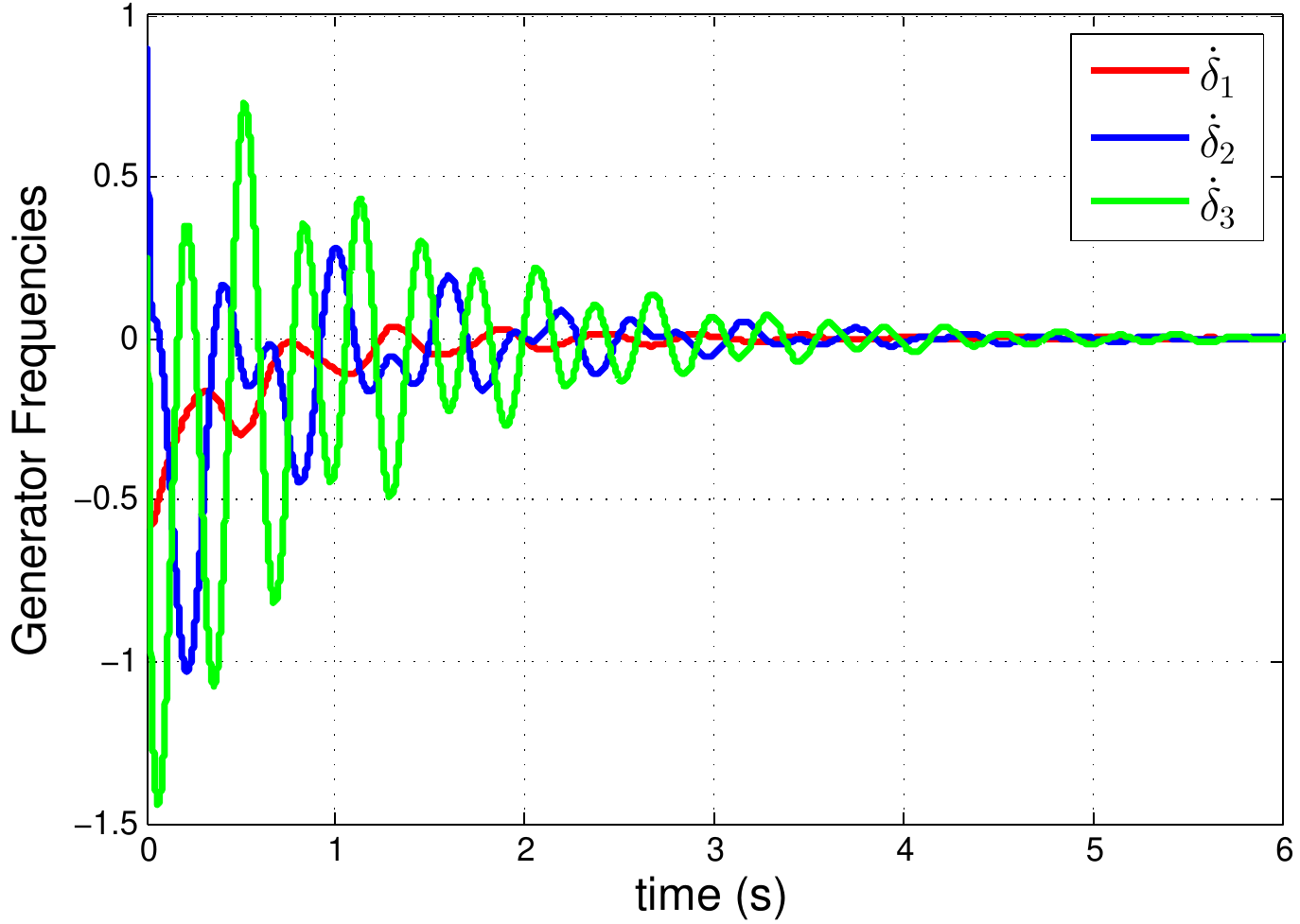}
\caption{Effect of power injection control: Convergence of generators frequencies to the base value.} 
\label{fig.InjectionControl_Frequency}
\end{figure}

Next, we apply the fault-dependent stability certificate in Section III.B. With the new equilibrium point $\delta_1^*$, we have a family of Lyapunov functions satisfying LMIs \eqref{eq.NewQKH}-\eqref{eq.NewQKH1}. Using the adaptation algorithm presented in \cite{VuTuritsyn:2014}, after some steps we find that there is a Lyapunov function in this family such that $V(x_0)<V_{\min}.$ As such, when we turn on the new power injections, the post-fault dynamics are stable and the post-fault trajectory will converge from the fault-cleared state $x_0$ to the new equilibrium point $\delta^*_1$. After that, we switch power injections back to the original values.

\subsubsection{Designing the other equilibrium points by changing transmission susceptances}

Using the adaptation algorithm, we do not find a suitable Lyapunov function certifying that $\delta_1^* \in {\bf SR_{origin}}$. As such, the
new equilibrium point $\delta_1^*$ may stay outside the stability region of the original equilibrium point $\delta^*_{\bf origin}$. We design the impedance adjustment controllers to render the post-fault dynamics from the new equilibrium point back to the original equilibrium point.

Assume that the impedances of transmission lines $\{1,4\}, \{2,7\}, \{3,9\}$ can be adjusted by FACTS devices integrated with these lines. The distance from the first equilibrium point to the original equilibrium point is calculated as $d_1(\delta^*_1,\delta^*_{\bf origin})=70.6424.$ Let $d = d_1(\delta^*_1,\delta^*_{\bf origin})/2 +1=36.3212,$ and solve the following convex QCQP with variable $B^{(2)}_{14}, B^{(2)}_{27},$ and $B^{(2)}_{39}:$ 
\begin{align}
\label{eq.DesignSecondEquilibrium}
    &\min_{\{B^{(2)}_{kj}\}} d_2(\delta^*_2,\delta^*_{1}) \\
    {\bf s.t.\;\; } & d_2(\delta^*_2,\delta^*_{\bf origin}) \le d_{1}(\delta^*_{1},\delta^*_{\bf origin})-d=  34.3212. \nonumber
\end{align}
Solving this convex QCQP problem, we obtain the new susceptances at transmission lines $\{1,4\},\{2,7\}, \{3,9\}$ as $B^{(2)}_{14}=33.4174 p.u.,
   B^{(2)}_{27}=22.1662 p.u.,$ and $B^{(2)}_{39}=   24.3839 p.u.,$ with which the distance from the second equilibrium point to the first equilibrium point and the original equilibrium point are given by $d_2(\delta^*_2,\delta^*_{1})= 60.9209$ and $d_2(\delta^*_2,\delta^*_{\bf origin})=34.3212.$ Using the adaptation algorithm,  we can check that $\delta^*_1 \in {\bf SR}_2$ and $\delta^*_2 \in {\bf SR}_{\bf origin}.$

\subsubsection{Simulation results}

When there is no control in use, the post-fault dynamics evolve as in Fig. \ref{fig.NoControl_Angle} in which we can see that the angle of the load bus 5 significantly deviates from that of other buses with the angular differences larger than 6. This implies that the post-fault dynamics evolve to a different equilibrium point instead of the desired stable equilibrium point $\delta_{\bf origin}^*,$ where the angular differences are all smaller than 0.6.

We subsequently perform the following control actions:
\begin{itemize}
\item [(i)] Changing the power injections of generators 1-3 and controllable load buses 4-6 to $P_1= 0.5890, P_2=
    0.5930, P_3=
    0.5989, P_4=
   -0.0333, P_5=
   -0.0617, P_6=
   -0.0165.$ From Fig. \ref{fig.InjectionControl_Angle} and Fig \ref{fig.InjectionControl_Distance}, it can be seen that the bus angles of the post-fault dynamics converge to the equilibrium point of the controlled post-fault dynamics which is the first equilibrium point $\delta_1^*.$ In Fig. \ref{fig.InjectionControl_Frequency}, we can see that the generator frequencies converge to the nominal frequency, implying that the post-fault dynamics converge to the stable equilibrium point  $\delta_1^*.$ However, it can be seen that the frequencies remarkably fluctuate. The fluctuation happens because we only change the power injection one time and let the post-fault dynamics automatically evolve to the designed equilibrium point $\delta_1^*.$ This is different from using the AGC where the fluctuation of the generator frequencies is minor, however we need to continuously measure the frequency and continuously update the control. 
\item [(ii)] To recover the resource spent for the power injection control, we switch the power injections to the original value. At the same time, we change the susceptances of transmission lines $\{1,4\}, \{2,7\},$ and $\{3,9\}$ to $B^{(2)}_{14}=33.4174 p.u.,
   B^{(2)}_{27}=22.1662 p.u.,$ and $B^{(2)}_{39}=   24.3839 p.u.$ The system trajectories will converge from the first equilibrium point $\delta_1^*$ to the second equilibrium point $\delta_2^*$, as shown in Figs. \ref{fig.ImpedanceControl_Angle}-\ref{fig.ImpedanceControl_Frequency}. Similar to the power injection control, in this case we also observe the fluctuation of generator frequencies, which is the result of the one-time change of line susceptances and autonomous post-fault dynamics after this change.

\begin{figure}[t!]
\centering
\includegraphics[width = 3.2in]{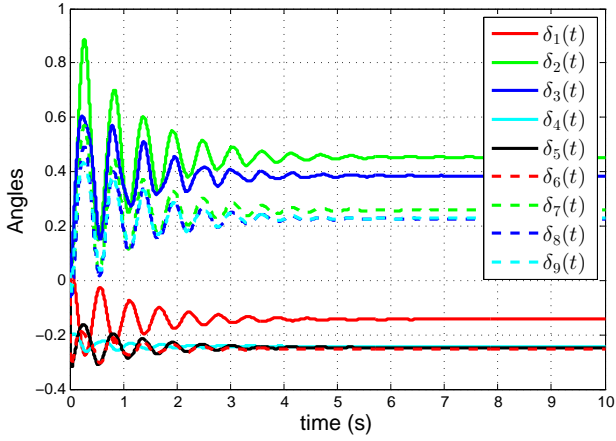}
\caption{Effect of susceptance control: Convergence of buses angles from $\delta_1^*$ to the second equilibrium point $\delta_2^*$ in post-fault dynamics.} \label{fig.ImpedanceControl_Angle}
\end{figure}

\begin{figure}[t!]
\centering
\includegraphics[width = 3.2in]{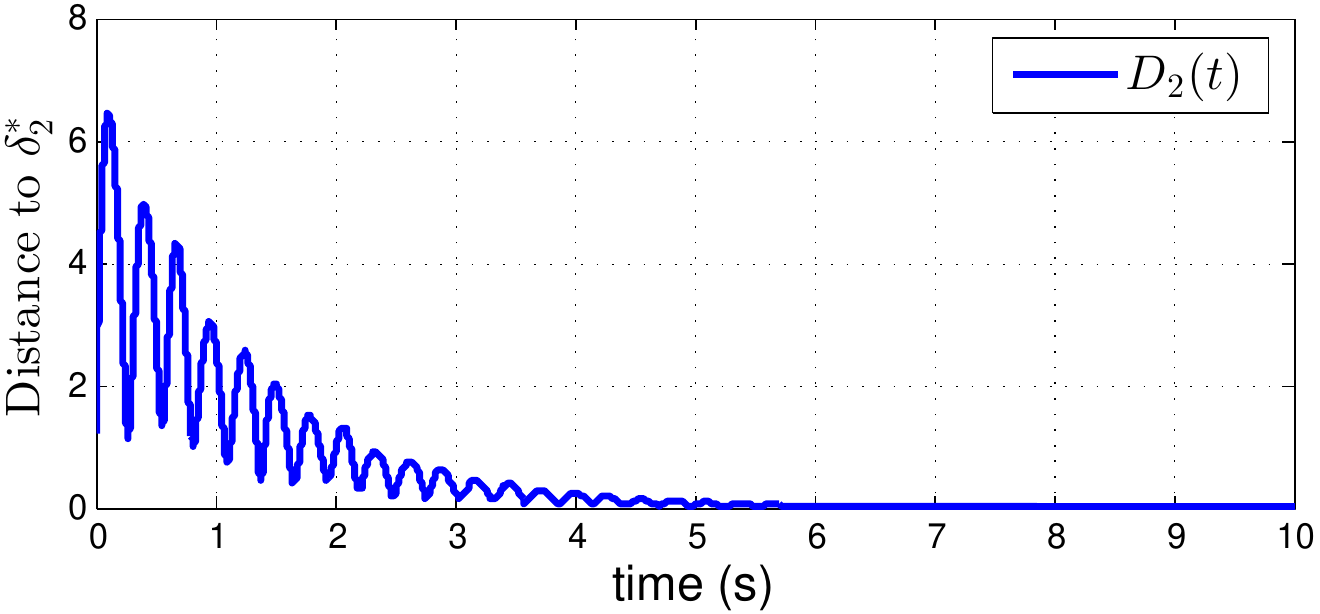}
\caption{Effect of susceptance control: the convergence of the distance $D_2(t)$ to $0$. Here, the Euclid distance $D_2(t)$ between a post-fault state and the second equilibrium point $\delta_2^*$
is defined as $D_2(t)=\sqrt{\sum_{i=2}^{9} (\delta_{i1}(t)-\delta_{2_{i1}}^*)^2}$.} \label{fig.ImpedanceControl_Distance}
\end{figure}

\begin{figure}[t!]
\centering
\includegraphics[width = 3.2in]{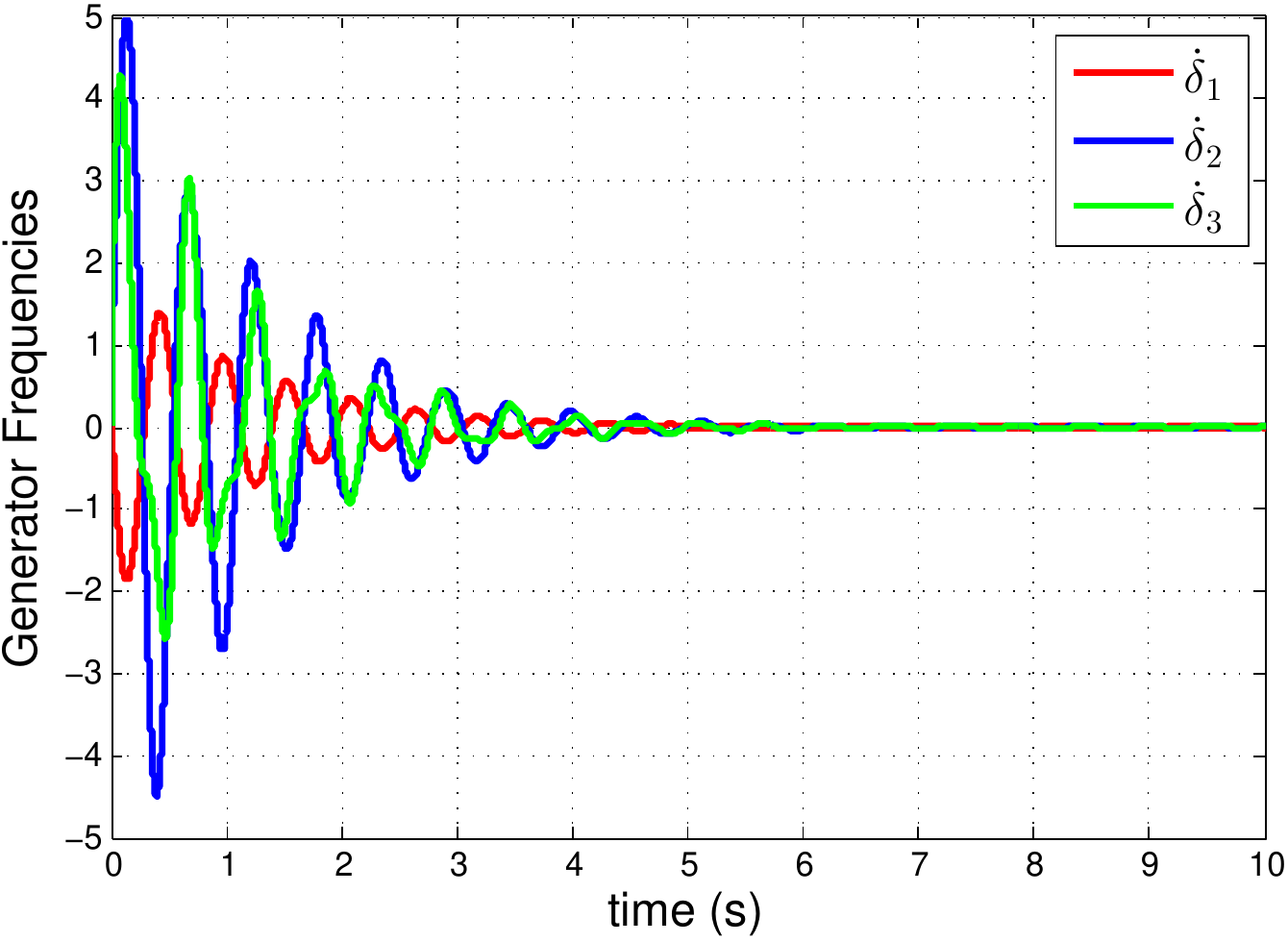}
\caption{Effect of susceptance control: Convergence of generators frequencies to the base value.} 
\label{fig.ImpedanceControl_Frequency}
\end{figure}

\item [(iii)] Switch the susceptances of transmission lines $\{1,4\}, \{2,7\},$ and $\{3,9\}$ to the original values. The system trajectories will autonomously converge from the second equilibrium point to the original equilibrium point $\delta^*_{\bf original},$ as shown in Fig. \ref{fig.Autonomous_Distance}.
\begin{figure}[t!]
\centering
\includegraphics[width = 3.2in]{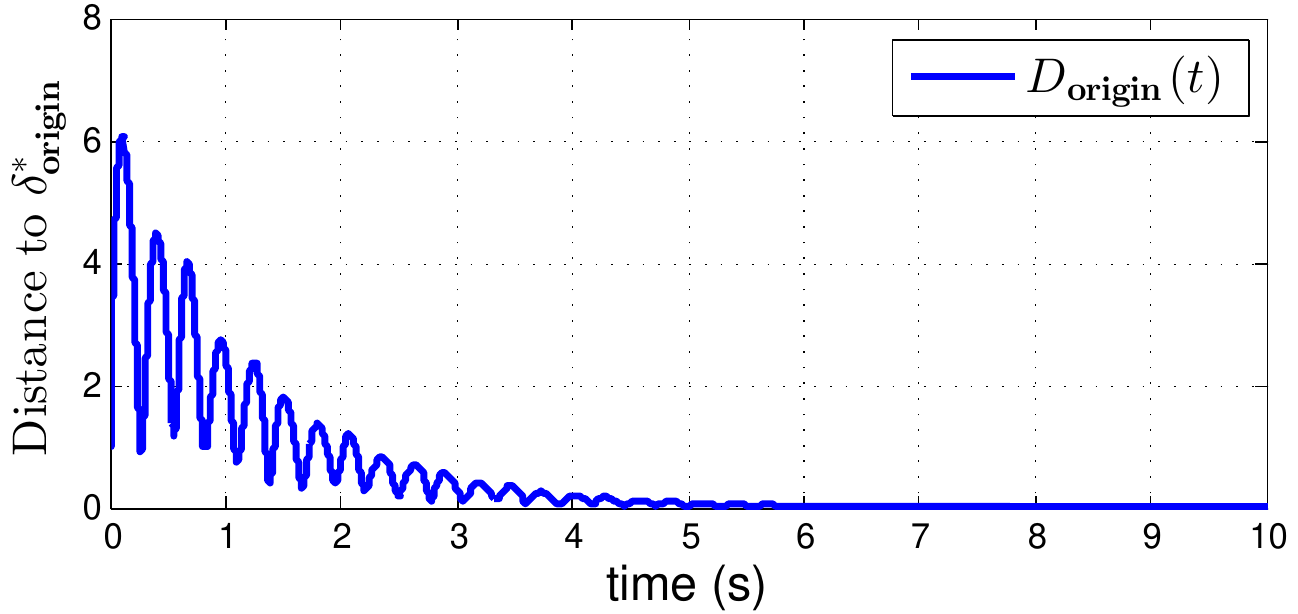}
\caption{Autonomous dynamics when we switch the line susceptances to the original values: the convergence of the distance $D_{\bf origin}(t)$ to $0$. Here, the Euclid distance $D_{\bf origin}(t)$ between a post-fault state and the original equilibrium point $\delta_{\bf origin}^*$
is defined as $D_{\bf origin}(t)=\sqrt{\sum_{i=2}^{9} (\delta_{i1}(t)-\delta_{{\bf origin}_{i1}}^*)^2}$.} \label{fig.Autonomous_Distance}
\end{figure}
\end{itemize}

\subsection{Scalability demonstration on 118 bus system}
The scalability of the proposed control design depends on minimizing $\|L^{\dag}p\|_{\mathcal{E},\infty}$ to find the optimum power injections $p^*$ and solving the quadratically
constrained quadratic program (QCQP) \eqref{eq.DefiningEquilibrium} to find the optimum line susceptances. Minimizing $\|L^{\dag}p\|_{\mathcal{E},\infty}$ is a linear problem and can be solved extremely fast even with the high number of variables. The QCQP \eqref{eq.DefiningEquilibrium} is a convex problem, and can also be solved quickly in large power systems if we have a small number of susceptance variables. 

\begin{figure}[t!]
\centering
\includegraphics[width = 3.2in]{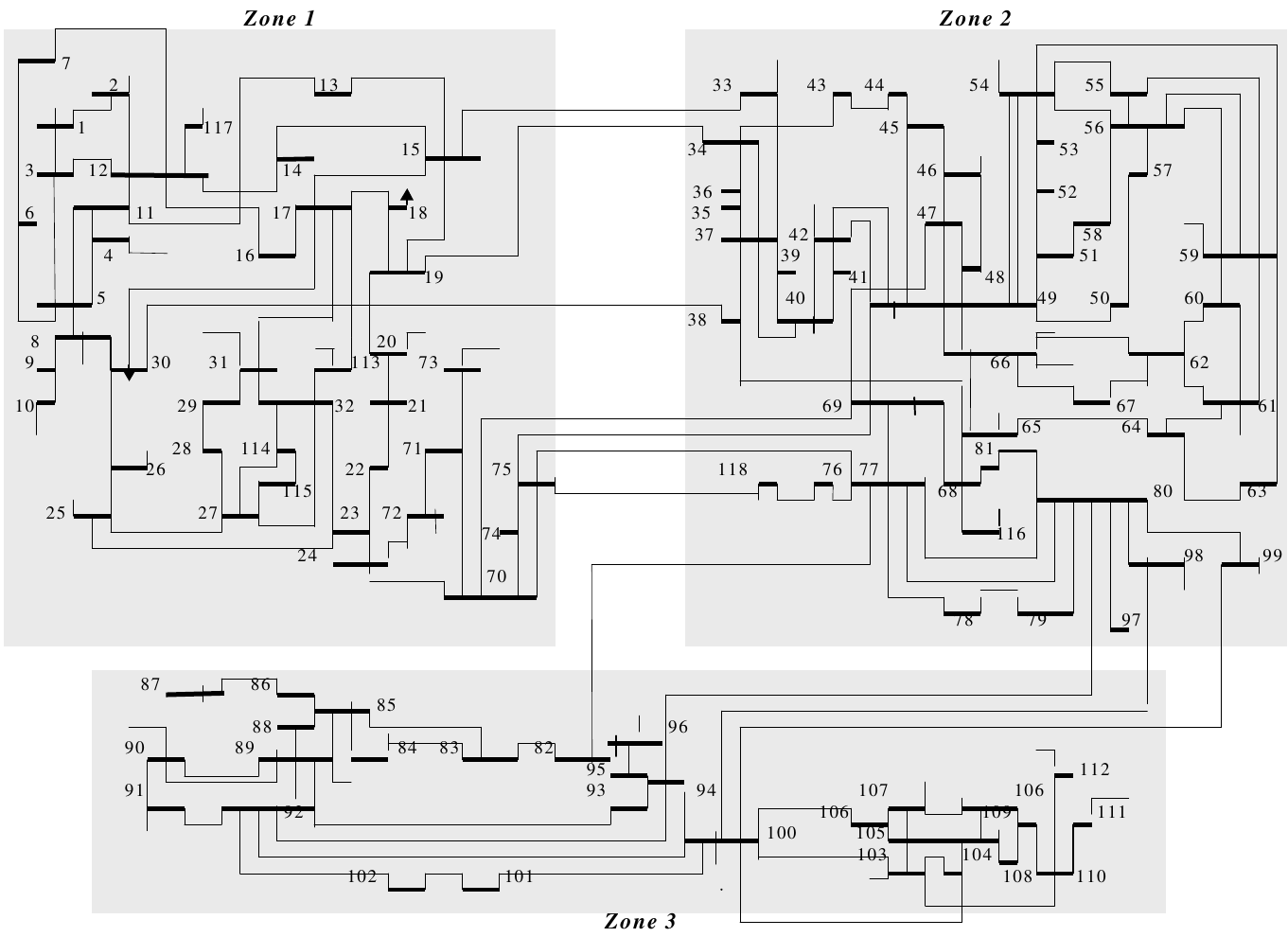}
\caption{IEEE 118-bus test case}
\label{fig.IEEE118}
\end{figure}

To clearly demonstrate the scalability of the proposed control method to large scale power systems, we utilize the  modified IEEE 118-bus test case \cite{118bus}, of which 54 are generator buses and the other 64 are load buses as showed in Fig. \ref{fig.IEEE118}. The data is taken directly from the test files \cite{118bus}, otherwise specified. The damping and inertia are not given in the test files and thus are randomly selected in the following ranges: $m_i \in [0.02,0.04], \forall i \in \mathcal{G} ,$ and $d_i\in [0.01,0.02], \forall i \in \mathcal{N}.$    
The grid originally contains 186 transmission lines. We eliminate 9 lines whose susceptance is zero, and combine 7 lines $\{42,49\},\{49,54\},\{56,59\},\{49,66\},\{77,80\},$ $\{89,90\},$ and $\{89,92\},$ each of which contains double transmission lines as in the test files \cite{118bus}.  Hence, the grid is reduced to 170 transmission lines connecting 118 buses. Assume that we can use the integrated FACTS devices to change the susceptances of the 3 transmission lines $\{19,34\}, \{69,70\},$ and $\{99,100\}$ which connect generators in different Zones 1, 2, and 3. These transmission lines may have strong effects on keeping the synchronization of the whole system.

\begin{figure}[t!]
\centering
\includegraphics[width = 3.2in]{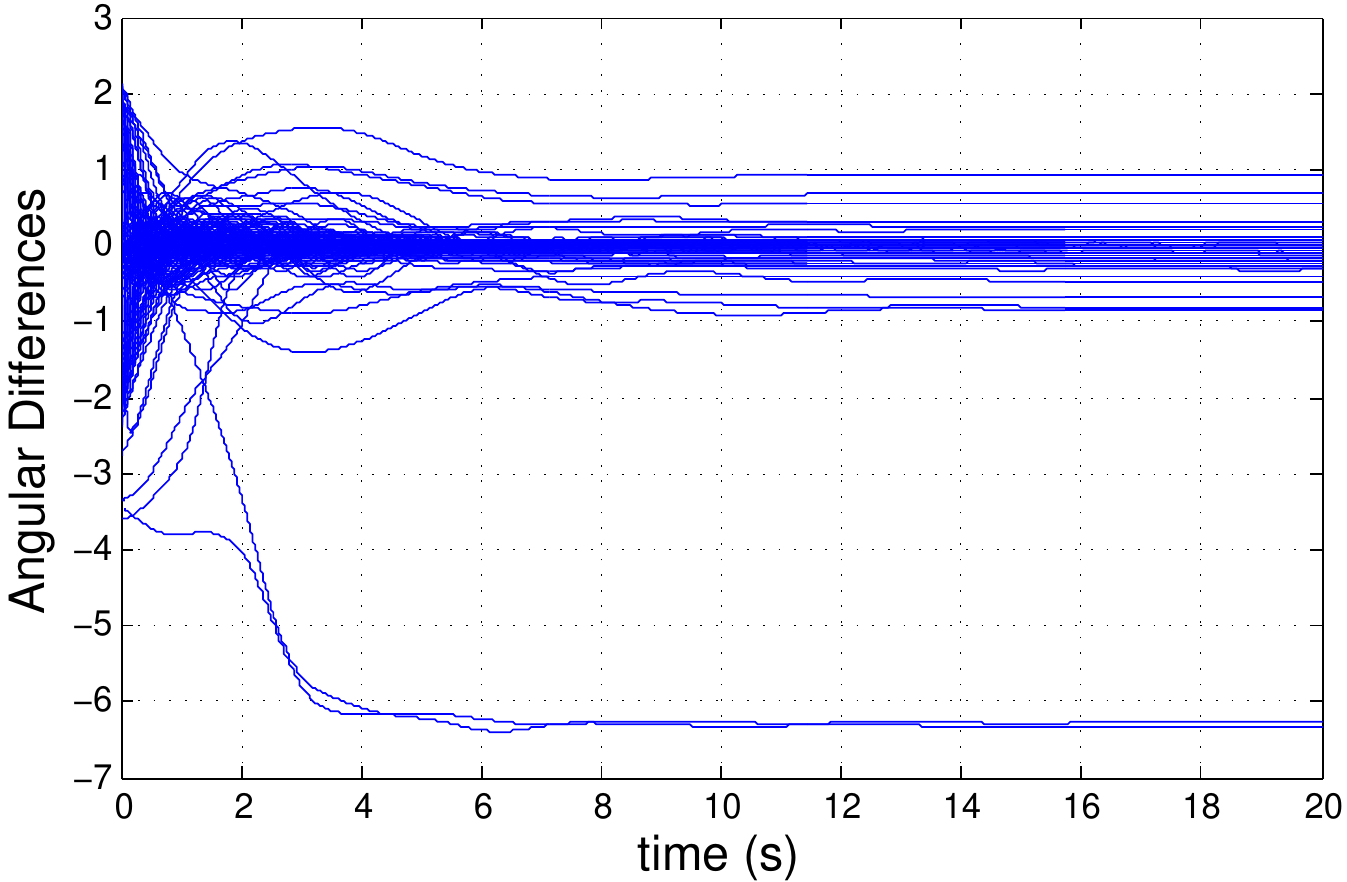}
\caption{Dynamics of buses angle differences in post-fault dynamics when the proposed control is not applied.} \label{fig.118bus_NoControl}
\end{figure}

We renumber the generator buses as $1-54$ and load buses as $55-118$. Assume that each of the first ten generator buses increases $0.01p.u.$ and each of the first ten load buses decreases $0.01p.u.,$ which result in an equilibrium point with $\|L^{\dag}p\|_{\mathcal{E},\infty}=0.8383.$ This equilibrium point stays near the stability margin $\delta_{kj}=\pi/2,$ and weakly stable. As a result, the fault-cleared state $\delta_{fault-cleared}$ does not stay inside the stability region of this equilibrium point, as can be seen from Fig. \ref{fig.118bus_NoControl} which shows that the uncontrolled post-fault dynamics converges to an equilibrium point with some angular differences larger than $\pi$. 

\begin{figure}[t!]
\centering
\includegraphics[width = 3.2in]{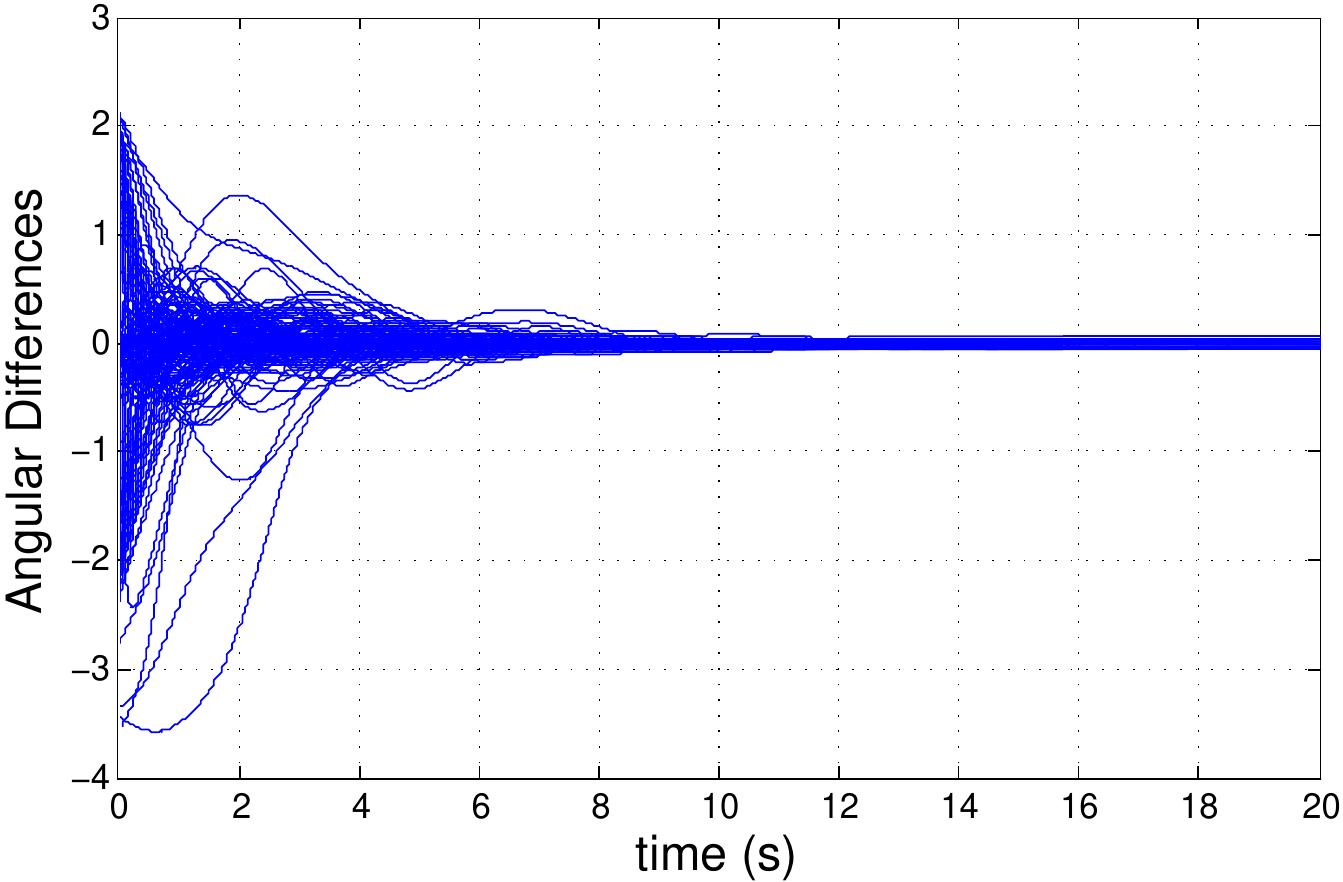}
\caption{Convergence of buses angle differences in post-fault dynamics under the control to the designed equilibrium point.} \label{fig.118bus_InjectionControl}
\end{figure}

Assume that we can control the power generation at generator buses $1-20,$ the load buses $55-64$ are deferrable, and the terminal loads at other buses are fixed. We design the first equilibrium point by changing the power injections of the generators 1-20 and load buses 55-64. Using CVX software to minimize $\|L^{\dag}p\|_{\mathcal{E},\infty},$ after less than 1 second, we obtain the optimum power injections at these controllable buses with
the minimum value of $\|L^{\dag}p\|_{\mathcal{E},\infty}=0.0569 < \emph{\emph{sin}}(\pi/55).$ Accordingly, the new equilibrium point $\delta_1^*$ is strongly stable since it stays far away from the stability margin $\delta_{kj}=\pi/2.$ The controlled post-fault dynamics converges from the fault-cleared state to the designed equilibrium point as showed in Fig. \ref{fig.118bus_InjectionControl}.

Now, we change the susceptances of the above selected transmission lines, which are $\{9,16\}, \{30,31\}$, and $\{44,45\}$ in the new order. Using CVX software in a normal laptop to solve the convex QCQP with variable set $\mathcal{B}=\{B^{(2)}_{\{9,16\}}>0, B^{(2)}_{\{30,31\}}>0,B^{(2)}_{\{44,45\}}>0 \},$ 
\begin{align}
    &\min_{\mathcal{B}} d_2(\delta^*_2,\delta^*_{1}) \\
    {\bf s.t.\;\; } & d_2(\delta^*_2,\delta^*_{\bf origin}) \le d_{1}(\delta^*_{1},\delta^*_{\bf origin})-0.001, \nonumber
\end{align}
 we obtain the optimum susceptances at transmission lines $\{9,16\}, \{30,31\}$, and $\{44,45\}$ in less than one second: 
 $B^{(2)}_{\{9,16\}}=0.0005p.u., B^{(2)}_{\{30,31\}}=0.0008p.u.,B^{(2)}_{\{44,45\}}=0.0012p.u.$. Therefore, the proposed control method can quickly determine the optimum values of both power injection and susceptance controls, and hence, it is suitable to handle faults in large scale power systems.  

\section{Conclusions}
\label{sec.conclusion}

This paper proposed a novel emergency control paradigm for power grids by exploiting the transmission facilities widely available on the grids. In particular, we formulated a control problem to recover the transient stability of power systems by adjusting the transmission susceptances of the post-fault dynamics such that a given fault-cleared state, that originally can lead to unstable dynamics, will be attracted to the post-fault equilibrium point. To solve this problem, we extended our recently introduced Lyapunov function family-based transient stability certificate \cite{VuTuritsyn:2014,VuTuritsyn:2015TAC} to a new set of convex fault-dependent functions. Applying this stability certificate, we determined suitable amount of transmission susceptance/power injection to be adjusted in critical/emergency situations.  We showed that the considered control design can be quickly performed through solving a number of linear and convex optimization problems in the form of SDP and convex QCQP. The advantage of the proposed control is that the transmission line's susceptance or power injection only needs to be adjusted one time in each step, and hence, no continuous measurement is required, as in the classical control setup. 


To make the proposed emergency control scheme applicable in practice, we need to take into account the computation and regulation delays, either by offline scanning contingencies and calculating the emergency actions before hand, or by allowing specific delayed time for computation. Also, the variations of load and generations during this delayed time should be considered. 
On the theoretical side, several questions are still open not only for power grids, but also for the general complex networks:
\begin{itemize}
    \item At which locations are the suitable transmission lines to adjust susceptances such that we can drive the post-fault dynamics from a given initial state to the desired equilibrium point?
    \item Given a grid, what is the minimum number of lines required to adjust susceptances to obtain the control objective? How many equilibrium points should be designed?
    \item What are the emergency situations where the proposed control scheme is not effective? Can the proposed control scheme be extended to deal with situations of voltage instability?
\end{itemize}

Finally, the installation of FACTS devices is certainly associated with non-negligible costs for the power system stakeholders. However, this paper does not advocate the installation of new FACTS devices solely for emergency control. It rather proposes the use of existing FACTS devices, e.g. PSTs, TCSCs, or HVDC, to assist in emergency control situations. For example, a large number of PSTs has been installed in several power systems for power flow control. HVDC lines and Back-to-Back converters become more and more widespread in systems in Europe, the US, or Asia.  In this paper, we propose to use only a number of these already installed devices, in order to ensure power system stability in emergency situations. The proposed method can also be combined with transmission line switching, an approach already used by operators to ensure power system security or minimize losses. This will however lead to a mixed-integer optimization problem, instead of the convex QCQP optimization problem as in Section IV.C of this paper. In that case, convex relaxations should be considered to make the control design computationally tractable \cite{6502760}.



\section{Appendix}

\subsection{Adaptation algorithm to find suitable Lyapunov function}
\label{appendix}

The family of Lyapunov functions characterized by the matrices $Q,K$ satisfying  LMIs \eqref{eq.NewQKH}-\eqref{eq.NewQKH1} allow us to find a
Lyapunov function that is best suited for a given fault-cleared state
$x_{0}$ or family of initial states. In the
following, we propose a simple algorithm  for the
adaptation of Lyapunov functions to a given initial state $x_0$ (similar to that in \cite{VuTuritsyn:2014}).

\vskip 0.2cm

\noindent Let $\epsilon$ be a positive constant.
\begin{itemize}
  \item[$-$] \emph{Step 1:} Find $Q^{(1)}, K^{(1)}$ by solving LMIs \eqref{eq.NewQKH}-\eqref{eq.NewQKH1}.
  Calculate $V^{(1)}(x_{0})$ and $V^{(1)}_{\min}$.
  \item[$-$] \emph{Step $k$:} If $x_{0} \notin
\mathcal{R}(Q^{(k-1)},K^{(k-1)}),$ (i.e., $V^{(k-1)}(x_{0})\ge V^{(k-1)}_{\min}$), then find matrices $Q^{(k)}, K^{(k)}$ by solving the
  following LMIs:
\begin{align}
    &\left[\begin{array}{ccccc}
          A^\top Q^{(k)}+Q^{(k)}A  & R \\
          R^\top  & -2H^{(k)}\\
        \end{array}\right] \le 0, \nonumber \\
& Q^{(k)}- \sum_{j\in \mathcal{N}_i}K^{(k)}_{\{i,j\}}C_{\{i,j\}}^\top C_{\{i,j\}} \ge 0, \nonumber\\
      & V^{(k)}(x_0) \le V^{(k-1)}_{\min}-\epsilon, \nonumber
\end{align}
with $R = Q{(k)}B-C^\top H{(k)}-(K{(k)}CA)^\top.$ Note that, $V^{(k)}(x_0)$ is a linear function of $Q^{(k)}, K^{(k)}.$ 
\end{itemize}

With this algorithm, we have
\begin{align}
 V^{(k-1)}_{\min} &\le V^{(k-1)}(x_{0}) \le V^{(k-2)}_{\min}-\epsilon
\le ... \le V^{(1)}_{\min}-(k-2)\epsilon.
\end{align}
Since $V^{(k-1)}_{\min}$ is lower bounded, this algorithm will terminate
after a finite number of the steps. There are two alternative
exits then. If $V^{(k)}(x_{0}) < V^{(k)}_{\min},$ then the Lyapunov
function is identified. Otherwise, the value of $\epsilon$ is
reduced by a factor of $2$ until a valid Lyapunov function is
found. Therefore, whenever the stability certificate of the given
initial condition exists, this algorithm possibly finds it after a
finite number of iterations.


\bibliographystyle{IEEEtran}
\bibliography{lff}
\end{document}